\documentclass[a4paper]{article}

\usepackage{INTERSPEECH2019}
\usepackage{amsmath,graphicx}
\usepackage{multirow}
\usepackage{mathtools,xparse}
\usepackage{graphicx}
\usepackage[dvips]{epsfig}
\usepackage{adjustbox}
\usepackage{subcaption}

\usepackage{cite}
\usepackage{array}
\usepackage{makecell}
\usepackage{multicol}

\usepackage{amsmath}
\usepackage{mathabx}
\usepackage{caption}
\usepackage{algorithm}
\usepackage{algpseudocode}

\usepackage{tabularx}
\usepackage{float}
\floatstyle{plaintop}
\restylefloat{table}
\usepackage[justification=centering]{caption}

\usepackage{subcaption}

\usepackage{chngpage}

\usepackage{array}
\newcolumntype{L}[1]{>{\raggedright\let\newline\\\arraybackslash\hspace{0pt}}m{#1}}
\newcolumntype{C}[1]{>{\centering\let\newline\\\arraybackslash\hspace{0pt}}m{#1}}
\newcolumntype{R}[1]{>{\raggedleft\let\newline\\\arraybackslash\hspace{0pt}}m{#1}}
\newcolumntype{?}{!{\vrule width 1pt}}

\title{Parameter enhancement for MELP speech codec in noisy communication environment}
\name{{Min-Jae Hwang$^{1, 2}$ and Hong-Goo Kang$^2$}
	\address{$^1$Search Solutions Inc., Seongnam, Korea\\
	$^2$Department of Electrical and Electronic Engineering, Yonsei University, Seoul, Korea}
	\email{hmj234@dsp.yonsei.ac.kr, hgkang@yonsei.ac.kr}}

\begin{document}
	\maketitle
	\begin{abstract}
	In this paper, we propose a deep learning (DL)-based parameter enhancement method for a mixed excitation linear prediction (MELP) speech codec in noisy communication environment.
	Unlike conventional speech enhancement modules that are designed to obtain clean speech signal by removing noise components before speech codec processing, the proposed method directly enhances codec parameters on either the encoder or decoder side.
	As the proposed method has been implemented by a small network without any additional processes required in conventional enhancement systems, e.g., time-frequency (T-F) analysis/synthesis modules, its computational complexity is very low.
	By enhancing the noise-corrupted codec parameters with the proposed DL framework, we achieved an enhancement system that is much simpler and faster than conventional T-F mask-based speech enhancement methods, while the quality of its performance remains similar.		
	\end{abstract}
	
	\section{Introduction}
	To build a comfortable voice communication system in noisy environment, it is necessary to include a speech enhancement or noise reduction techniques \cite{Kleijn:1995:SCS:546671, collura1999melp, agarwal2002preprocessing, rainer2004new}.
	However, the core module of coding system, i.e., \textit{vocoding} techniques, and the speech enhancement techniques have been developed independently to each other.
	Thus, the entire speech communication system is implemented by simply concatenating the two systems, that is, the enhancement module is processed first, and the vocoder is processed afterwards
	In this method, the characteristics of the coding process are not adequately considered during the enhancement process, and it is difficult to include conventional enhancement systems in the speech coding system with their maximum performance, especially due to the additional but necessary processes required in the speech enhancement system.
	
	Although the cost for modernized communication devices has dropped significantly thanks to advances in semiconductor technologies, it is still an important issue in many underdeveloped countries, where low-cost processors operating in a limited communication bandwidth environment are common.
	Thus, the need for an effective low bit-rate speech coding technique is still very high.
	Based on the assumption that we can modify the core algorithm of the conventional low bit-rate speech coding standard, we chose the 2.4 kbit/s mixed excitation linear prediction (MELP) speech coder as our target system \cite{mccree1995mixed, Supplee1997MELP}.
	Several statistical-based speech enhancement techniques were combined with the MELP codec to improve the robustness of the codec in background noise conditions \cite{collura1999melp, agarwal2002preprocessing}, but the improvement was not dramatic especially in non-stationary noise environment.
	
	To improve the quality of speech enhancement systems, recently developed deep learning (DL) techniques have been actively utilized \cite{narayanan2013ideal, erdogan2015phase, xugang2014speech}.
	Typical examples are time-frequency (T-F) masking-based algorithms \cite{narayanan2013ideal, erdogan2015phase}. 
	These algorithms first determine target T-F masking values with various signal-to-noise ratio (SNR) criteria, then they train a DL network using a noisy spectrum input and the output of target masking values in each T-F bin. 

	However, these approaches operate at the frequency domain and require the use of Fourier transform and overlap-add (OLA) processes; thus, additional delay and computational complexity is unavoidable, which makes them unsuitable for a low bit-rate coding system. 
	Moreover, the approach is not a good choice in terms of memory usage, because the large dimensions of the input/output of the DL network corresponding to the frequency domain resolution require the use of a large DL network.
	Although several algorithms directly operate in the time-domain speech signal have been proposed \cite{Stoller2018wave, Pascual2017SEGANSE, Rethage2018Wavenet}, they are unsuitable for voice communication systems, because they are typically designed to have a non-causal structure with rich resources.
	
	In this paper, we propose a DL-based vocoder parameter enhancement system that can be directly applied to the 2.4 kbit/s MELP-based speech communication system.
	Given the coding parameters encoded by the MELP encoder in noisy environment, e.g., line spectral frequencies (LSFs), gain, pitch, or other excitation related parameters, we use them as an input to the DL network, then directly estimate the parameters as they are obtained from the clean speech input. 
	As the proposed enhancement works in vocoder parameterization/waveform reconstruction processes and does not require any T-F analysis/synthesis process, its complexity is very low, and there is no additional delay. 
	In addition, as the dimension of input and output features is small, it is possible to build a very small-size DL network; we only needed 182.11 KB memory, even with a 32-bit single-precision floating point format.
	The objective and subjective test results confirm that the proposed vocoder parameter enhancement algorithm provides much simpler and faster enhancement architectures than the conventional speech enhancement systems while retaining a high quality.
	
	\section{MELP coder with speech enhancement}
	\label{sec:background}
	\subsection{MELP vocoder}
	\label{sec:melp_vocoder}
	The main characteristic of the MELP codec \cite{mccree1995mixed} is to model an excitation signal by mixing voiced pulse and noise components in the frequency domain, where bandpass voicing flags are used to represent the voicing information of frequency subbands.
	In the system, total six parameters that consist of excitation and spectral parameters are transmitted to the decoder.
	Spectral parameters are represented by 10-dimensional line spectral frequencies (LSFs) and two dimensional gain terms, whereas excitation parameters consist of 5-dimensional bandpass voicing flags, pitch value, aperiodicity flag, and 10-dimensional Fourier magnitudes at a frame rate of 22.5-ms.
	
	To obtain the voiced excitation component, the first 10 harmonics of the impulse train with an interval of pitch period is shaped with the decoded Fourier magnitudes.
	If the aperiodicity flag is on, then the jitter effect is imposed to the pulse excitation by randomly shifting the position of impulse component within a 25\% range of the pitch value. 
	Then, the mixed excitation signal is obtained by summing the pulse and noise excitation components after applying the bandpass filters processed by following the bandpass voicing flags. 
	Finally, the speech signal is reconstructed by applying a linear prediction synthesis process.
	To make the reconstructed speech signal sound natural, the adaptive spectral enhancement filter and pulse dispersion filter are applied.
	
	\subsection{Speech enhancement as pre-processing algorithm}
	\label{sec:conventional_speech_enhance}
\begin{figure}[!t]
		\centering
		\hspace*{-4mm}
		\begin{tabular}{c}
			\begin{subfigure}[b]{0.45\textwidth}
				\centering
				\includegraphics[width=\linewidth]{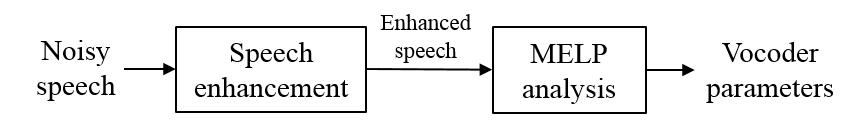}
				\caption{Pre-processing speech enhancement.}
			\end{subfigure}\vspace*{0.cm}
			\\ 
			\begin{subfigure}[b]{0.45\textwidth}
				\centering
				\includegraphics[width=\linewidth]{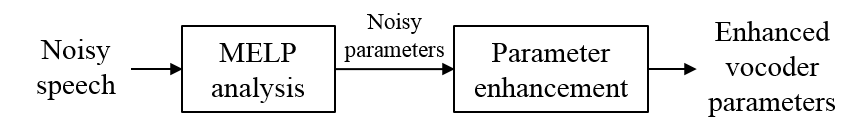}
				\caption{Encoder parameter enhancement.}
			\end{subfigure}\vspace*{0.0cm}
			\\
			\begin{subfigure}[b]{0.45\textwidth}
				\centering
				\includegraphics[width=\linewidth]{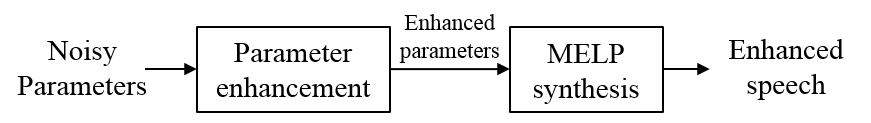}
				\caption{Decoder parameter enhancement.}
			\end{subfigure}\vspace*{0.0cm}
		\end{tabular}
		\caption{Various types of speech enhancement processes with MELP coder.}
		\label{fig:speech_enhance}
	\end{figure}

	To improve the quality of MELP coding system, the speech enhancement algorithms have been introduced as a pre-processing module of the MELP encoder as illustrated in the Fig.~\ref{fig:speech_enhance}-(a).
	For instance, Martin \textit{et al}. \cite{rainer2004new} introduced a minimum mean square error log spectral amplitude estimator (MMSE-LSA) speech enhancement system \cite{ephraim1985speech} as a pre-processing module.
	However, its performance was not satisfactory, especially in a non-stationary noise environment.
	
	By applying DL-based enhancement techniques, the quality of coding system can be further improved.
	In general, DL techniques are utilized to model the \textit{features} obtained by the power spectrum of clean and noisy speech.
	For instance, a widely used ideal ratio mask (IRM)-based DL network generates IRM components, which are the T-F masking components estimated by the function of SNR between clean and noise power spectrums.
	Then, the enhanced magnitude spectrum is obtained by multiplying the IRM components to the input noisy spectrum.
	Finally, a speech signal is reconstructed by applying an inverse Fourier transform and an OLA process.
	
	In spite of the high performance of conventional DL-based speech enhancement algorithm, it has several limitations to be used as a pre-processing module for a low bit-rate speech codec because of complexity, memory, and delay issues.
	For instance, the forward and inverse Fourier transforms, and OLA process required in the enhancement process present unavoidable delay and computational bottlenecks. 
	Moreover, the conventional enhancement algorithm requires a large DL network, because the dimension of input and output vectors is equal to the resolution of the Fourier analysis, which must be large to achieve high performance. 
	Thus, it is difficult to train a network with a small number of parameters.
	
	In the next section, we propose a parameter enhancement method for the 2.4 kbit/s MELP communication system, which is implementable by a very simple system architecture with low complexity, low additional memory usage, and no delay.

	\begin{table*}[t!]
		\caption{Objective evaluation results for the MELP codec system with various speech enhancement methods.
			In each of large and small networks, the system with better performance is represented in bold typeface.
			The labels `No enhance.', `IRM', `Param-Enc' and `Param-Dec' imply the MELP codec system without any enhancement module, with IRM-based enhancement, and with proposed parameter enhancement in encoder side and decoder side, respectively.}
		\label{table:obj_results}
		\setlength\tabcolsep{4.5pt} 
		\footnotesize
		
		\begin{subtable}{\linewidth}
			\centering
			\hspace*{-2mm}
			\begin{tabular}{c|c||c|c|c|c|c|c|c|c||c|c|c|c|c|c|c|c}
				\Xhline{2\arrayrulewidth}
				\multirow{4}{*}{\begin{tabular}[c]{@{}c@{}}\footnotesize Noise\\ \footnotesize type\end{tabular}} & \multirow{4}{*}{\begin{tabular}[c]{@{}c@{}}\footnotesize SNR \\ (dB)\end{tabular}}  & \multicolumn{8}{c||}{VUV error (\%)}                                                                                                                       & \multicolumn{8}{c}{Gain-RMSE}                                                                                                                            \\ \cline{3-18} 
				&                       & \multicolumn{2}{c|}{No enhance.}                        & \multicolumn{3}{c|}{Large network}                    & \multicolumn{3}{c||}{Small network}                    & \multicolumn{2}{c|}{No enhance.}                        & \multicolumn{3}{c|}{Large network}                    & \multicolumn{3}{c}{Small network}                    \\ \cline{3-18} 
				&  & \multirow{2}{*}{Clean} & \multirow{2}{*}{Noisy} & \multirow{2}{*}{IRM} & \multicolumn{2}{c|}{Param} & \multirow{2}{*}{IRM} & \multicolumn{2}{c||}{Param} & \multirow{2}{*}{Clean} & \multirow{2}{*}{Noisy} & \multirow{2}{*}{IRM} & \multicolumn{2}{c|}{Param} & \multirow{2}{*}{IRM} & \multicolumn{2}{c}{Param} \\ \cline{6-7} \cline{9-10} \cline{14-15} \cline{17-18} 
				&                       &                        &                        &                      & Enc         & Dec         &                      & Enc         & Dec         &                        &                        &                      & Enc         & Dec         &                      & Enc         & Dec         \\ \hline\hline
				\multirow{3}{*}{\footnotesize Seen}                                                 
				& 0                     & \multirow{6}{*}{7.31}  & 38.48                  & 23.24                & \textbf{15.87}        & 16.37        & 26.95                & \textbf{16.29}        & 16.74        & \multirow{6}{*}{3.24}  & 18.88                  & 8.48                 & \textbf{7.89}         & 8.10         & 8.45                 & \textbf{7.99}         & 8.14         \\ 
				& 5                     &                        & 29.52                  & 19.81                & \textbf{14.00}        & 15.05        & 21.57                & \textbf{14.49}        & 15.33        &                        & 15.82                  & 7.15                 & \textbf{6.60}         & 6.87         & 7.20                 & \textbf{6.74}         & 6.91         \\ 
				& 10                    &                        & 23.44                  & 16.85                & \textbf{12.84}        & 14.05        & 17.82                & \textbf{13.19}        & 14.00        &                        & 13.18                  & 6.14                 & \textbf{5.71}         & 5.99         & 6.20                 & \textbf{5.94}         & 5.96         \\  \cline{1-2} \cline{4-10} \cline{12-18} 
				\multirow{3}{*}{\footnotesize Unseen}                                               
				& 0                     &                        & 42.83                  & 27.34                & \textbf{18.25}        & 19.24        & 31.61                & \textbf{18.47}        & 19.24        &                        & 20.46                  & 11.09                & \textbf{9.82}         & 10.19        & 11.97                & \textbf{9.95}         & 10.19        \\ 
				& 5                     &                        & 33.01                  & 22.08                & \textbf{16.32}        & 16.82        & 24.55                & \textbf{16.07}        & 16.82        &                        & 17.40                  & 8.85                 & 8.19         & \textbf{8.16}         & 10.00                & \textbf{8.01}         & 11.55        \\ 
				& 10                    &                        & 24.83                  & 18.70                & \textbf{14.06}        & 15.06        & 20.05                & \textbf{14.03}        & 15.06        &                        & 14.56                  & 7.31                 & \textbf{6.52}         & 6.67         & 8.29                 & \textbf{6.71}         & 10.78        \\ \Xhline{2\arrayrulewidth}
			\end{tabular}
		\end{subtable}
		
		\begin{subtable}{\linewidth}
			\centering
			\vspace*{1mm}
			\hspace*{-3mm}
			\begin{tabular}{c|c||c|c|c|c|c|c|c|c||c|c|C{7mm}|c|C{7.5mm}|C{7mm}|c|C{7mm}}
				\Xhline{2\arrayrulewidth}
				\multirow{4}{*}{\begin{tabular}[c]{@{}c@{}}\footnotesize Noise\\ \footnotesize type\end{tabular}} & \multirow{2}{*}{SNR}  & \multicolumn{8}{c||}{F0-RMSE (Hz)}                                                                                                                         & \multicolumn{8}{c}{LSD (dB)}                                                                                                                             \\ \cline{3-18} 
				&                       & \multicolumn{2}{c|}{No enhance.}                        & \multicolumn{3}{c|}{Large network}                    & \multicolumn{3}{c||}{Small network}                    & \multicolumn{2}{c|}{No enhance.}                        & \multicolumn{3}{c|}{Large network}                    & \multicolumn{3}{c}{Small network}                    \\ \cline{2-18} 
				& \multirow{2}{*}{(dB)} & \multirow{2}{*}{Clean} & \multirow{2}{*}{Noisy} & \multirow{2}{*}{IRM} & \multicolumn{2}{c|}{Param} & \multirow{2}{*}{IRM} & \multicolumn{2}{c||}{Param} & \multirow{2}{*}{Clean} & \multirow{2}{*}{Noisy} & \multirow{2}{*}{IRM} & \multicolumn{2}{c|}{Param} & \multirow{2}{*}{IRM} & \multicolumn{2}{c}{Param} \\ \cline{6-7} \cline{9-10} \cline{14-15} \cline{17-18} 
				&                       &                        &                        &                      & Enc         & Dec         &                      & Enc         & Dec         &                        &                        &                      & Enc         & Dec         &                      & Enc         & Dec         \\ \hline\hline
				\multirow{3}{*}{\footnotesize Seen}                                                 
				& 0                     & \multirow{6}{*}{9.61}  & 14.32                  & 13.03                & \textbf{11.80}        & 11.82        & 12.98                & \textbf{11.99}        & 12.27        & \multirow{6}{*}{2.20}  & 6.22                   & 4.96                 & 4.69         & \textbf{4.66}         & 5.02                 & 4.76         & \textbf{4.75}         \\ 
				& 5                     &                        & 11.46                  & 11.83                & \textbf{11.04}        & 11.41        & 11.98                & \textbf{11.51}        & 11.53        &                        & 5.54                   & 4.33                 & 4.13         & \textbf{4.10}         & 4.39                 & 4.21         & \textbf{4.20}         \\ 
				& 10                    &                        & 10.66                  & 10.55                & \textbf{11.07}        & 11.18        & 11.14                & \textbf{10.69}        & 11.15        &                        & 4.86                   & 3.87                 & 3.71         & \textbf{3.67}         & 3.93                 & 3.80         & \textbf{3.78}         \\ \cline{1-2} \cline{4-10} \cline{12-18} 
				\multirow{3}{*}{\footnotesize Unseen}                                               
				& 0                     &                        & 12.89                  & 13.02                & \textbf{11.95}        & 12.84        & 13.87                & \textbf{12.47}        & 12.84        &                        & 6.97                   & 6.88                 & 5.87         & \textbf{5.83}         & 6.96                 & \textbf{5.77}         & 5.84         \\ 
				& 5                     &                        & 9.67                   & \textbf{10.76}       & 11.32        		 & 11.55        & 12.01                & \textbf{11.19}        & 11.55        &                        & 6.25                   & 5.84                 & 5.11         & \textbf{5.10}         & 5.92                 & \textbf{5.06}         & 5.10         \\ 
				& 10                    &                        & 10.19                  & 10.93                & \textbf{10.75}        & 10.78        & 10.88                & 11.14        & \textbf{10.78}        &                        & 5.43                   & 5.03                 & \textbf{4.42}         & 4.48         & 5.09                 & \textbf{4.47}         & 4.48         \\ \Xhline{2\arrayrulewidth}
			\end{tabular}
		\end{subtable}		
	\end{table*}

	\section{Vocoder parameter enhancement method}
	\label{sec:proposed}
	In the proposed system, the noise-corrupted speech signal is first parameterized to the MELP parameters without any pre-processing.
	Then, noisy MELP parameters are directly enhanced to be similar to the ones obtained from a clean speech signal via a DL network.
	To train the network, first, both noisy and clean MELP parameters are first extracted from the noisy and clean speech pair through MELP vocoder analysis as described in Section~\ref{sec:melp_vocoder}.
	Then, the DL network is trained to estimate clean MELP parameters from noisy MELP parameters by minimizing the mean squared error (MSE) criterion.
	
	To model the MELP parameters more accurately, some MELP parameters, such as gain, pitch, and Fourier magnitudes are refined before training the network.
	First, the mean and difference components of two dimensional gain features are used instead of the original gain features.
	Gain features imply the energy of two adjacent subframes, so independently generating gain features results in a discontinuous energy contour with a shimmer-like sound artifact.
	By coupling the two gain terms with a mean-difference pair, this problem can be easily solved.
	Moreover, because pitch values at the unvoiced region are set to zero at first time, discontinuity at the voice/unvoice boundary decreases the regression accuracy of pitch contour.
	We linearly interpolate the value of unvoiced region to prevent discontinuity at the voiced/unvoiced boundary.	
	Finally, we use logarithmic Fourier magnitudes to estimate their trajectory accurately.
	
	Since the MELP parameters are available in both the encoder and decoder step, we classify the proposed parameter enhancement method depending on the enhancement location to make the system have a higher flexibility as illustrated in Fig.~\ref{fig:speech_enhance}-(b) and (c), i.e., the \textbf{\textit{encoder side}} and \textbf{\textit{decoder side}} approaches, respectively.
	In the encoder side approach, noisy MELP parameters are first extracted from the noisy speech signal without any pre-enhancement processing, and the parameters are enhanced by the pre-trained DL network.
	Then, the enhanced MELP parameters are transmitted to the decoder after performing quantization and bitstream formatting processes.
	In the decoder side approach, on the other hand, the reconstructed MELP parameters at the decoder step are enhanced via pre-trained DL network. 
	Then, the MELP synthesis module reconstructs the speech signal through the use of enhanced MELP parameters.
		
	Note that the parameter enhancement approaches do not require any framing process, Fourier transform, or the heuristic selection of hyper-parameters such as the type of analysis window, window length, frame rate, or resolution of Fourier transform.
	Thus, we are able to build a simple and compact speech enhancement system that is fully specialized to the codec specifications.
	Its computational complexity is very low and there's no additional delay in the enhancement process.
	In addition, the total number of the DL network's input and output vectors is only 29-dimension; thus, their behavior is very simple in comparison to the features used in conventional speech enhancement, which enables us to train them successfully with a small size DL network.
		
	\section{Experiments}
	\label{sec:experiment}
	\subsection{Database generation}
	In the experiments, phonetically balanced TIMIT corpus \cite{timit} and NOISEX-92 corpus \cite{VARGA1993247} were used as speech and noise databases, respectively.
	To match the sampling rate with the 2.4 kbit/s MELP codec, all samples were down-sampled to 8-kHz.
	In the TIMIT database, sentences ``SA1'' and ``SA2'' commonly recorded by all speakers were excluded from the experiments; thus, a total of 3,696 utterances were used for training set, 1,152 utterances were used for validation set, and 48 utterances were used for test set.
	In the noise database, four types of noise; babble, factory1, volvo, and white noise, were used in a seen condition test, and other two types of noise; destroyer engine and pink noise, were used in an unseen condition test.
	To construct a noisy speech database for the training and validation processes, the speech signal was replicated and mixed with the noise signal under six SNR conditions from -5 dB to 20 dB with a step size of 5 dB.
	As a result, total 88,704 utterances (about 75 hours) and 22,648 utterances (about 24 hours) were used as a training and validation set, respectively.
	For testing purposes, the speech signal was mixed with both seen and unseen noises under four SNR conditions from 0 dB to 15 dB with a step size of 5 dB so that the 768 utterances for seen condition and 384 utterances for unseen condition were tested.
		\vspace{-2mm}
	\subsection{Network architectures}
	To show the performance of proposed systems, we included the MELP codec system with IRM-based speech enhancement algorithm \cite{narayanan2013ideal} as a baseline system.
	In each system, we tested various gated recurrent unit (GRU)-based DL networks \cite{cho-al-emnlp14} by varying the size; A \textbf{\textit{large network}} and a \textbf{\textit{small network}} were used to simulate the enhancement scenario with rich and limited resources, respectively.
	The detailed settings of each system are described below.

	\begin{table}[t!]
		\caption{STOI evaluation results for the MELP codec system with various speech enhancement methods.}
		\setlength\tabcolsep{3.3pt} 
		\hspace{-2mm}
		\label{table:stoi_results}
		\footnotesize
		\begin{tabular}{c|c||c|c|c|c|c|c|c|c}
			\Xhline{2\arrayrulewidth}
			\multirow{3}{*}{\begin{tabular}[c]{@{}c@{}} \footnotesize Noise\\ \footnotesize type\end{tabular}} & \multirow{2}{*}{SNR} & \multicolumn{2}{c|}{No enhance.}                        & \multicolumn{3}{c|}{Large network}                    & \multicolumn{3}{c}{Small network}                    \\ \cline{3-10} 
			&                      & \multirow{2}{*}{Clean} & \multirow{2}{*}{Noisy} & \multirow{2}{*}{IRM} & \multicolumn{2}{c|}{Param} & \multirow{2}{*}{IRM} & \multicolumn{2}{c}{Param} \\ \cline{2-2} \cline{6-7} \cline{9-10} 
			& (dB)                 &                        &                        &                      & Enc         & Dec         &                      & Enc         & Dec         \\ \hline\hline
			\multirow{3}{*}{\footnotesize Seen}                                                 
			& 0                    & \multirow{6}{*}{0.86}  & 0.62                   & \textbf{0.74}        & 0.72         & 0.71         & \textbf{0.72}        & 0.70         & 0.71         \\ 
			& 5                    &                        & 0.71                   & \textbf{0.80}        & 0.78         & 0.78         & \textbf{0.78}        & 0.77         & 0.77         \\ 
			& 10                   &                        & 0.79                   & \textbf{0.84}        & 0.82         & 0.82         & \textbf{0.83}        & 0.81         & 0.81         \\ \cline{1-2} \cline{4-10} 
			\multirow{3}{*}{\footnotesize Unseen}                                               
			& 0                    &                        & 0.56                   & \textbf{0.69}                 & 0.65         & 0.63         & \textbf{0.65}                 & 0.63         & 0.63         \\ 
			& 5                    &                        & 0.68                   & \textbf{0.78}                 & 0.75         & 0.74         & \textbf{0.75}                 & 0.73         & 0.74         \\ 
			& 10                   &                        & 0.77                   & \textbf{0.83}                 & 0.80         & 0.83         & \textbf{0.81}                 & 0.79         & 0.80         \\ \Xhline{2\arrayrulewidth}
		\end{tabular}
	\end{table}

	\vspace{-3mm}
	\subsubsection{Parameter enhancement system}	
	The MELP parameters refined by the method described in Section~\ref{sec:proposed} from the noisy-clean speech pair were composed of 29-dimensional input and output vectors of parameter enhancement.
	Before training, both the input and output features were normalized to have zero-mean and unit-variance.
	
	In the large network, hidden layers consisted of two GRU layers with 512 units at the input side and two feed-forward (FF) layers with 1,024 hidden nodes at the output side.
	In the small network, the hidden layers consisted of one GRU layer with 64 units at the input side and two FF layers with 128 hidden nodes at the output side.
	In total, the large network had around 4.01 million parameters which correspond to the 15.30 MByte and the small network had 46.62 thousand parameters which correspond to the 0.18 MByte, both with a 32-bit single-precision floating point.
	The rectified linear unit (ReLu) and linear activation functions were used in hidden and output layers, respectively.
	The weights were initialized using \textit{Xavier initializer} \cite{xavier2010init}, and trained using \textit{back-propagation through time} procedure with \textit{Adam} optimizer \cite{williams1990efficient, diederik2014adam} to optimize MSE criterion.
	
	In the enhancement step, the inverse processes of feature refinements were performed. 
	First, two gain features were reconstructed by summing and subtracting the mean and difference gain terms, respectively.
	Then, the pitch of unvoiced region was set to zero, and the exponential operator was performed to the Fourier magnitudes.
	Finally, the speech waveform was reconstructed using the MELP synthesizer.
	
	\vspace{-2mm}
		
	\subsubsection{IRM-based speech enhancement system}
	To assess system performance in fair condition, the size of the IRM-based DL network was set to similar to that of the parameter enhancement system.
	To compose input and output vectors, 129-dimensional log-power spectrum and IRM components were extracted from a 32-ms speech frame at a 22.5-ms frame rate with 9.5-ms overlap.
	In the large network, hidden layers consisted of two GRU layers with 512 units at the input side and two FF layers with 1,024 hidden nodes at the output side.
	In the small network, hidden layers consisted of one GRU layer with 64 units at the input side and two FF layers with 64 hidden nodes at the output side.
	In total, the networks had around 16.28 MByte and 0.21 MByte with a 32-bit single-precision floating point, respectively.
	The ReLu and sigmoid activation functions were used in the hidden and output layers, respectively.
	The weight initialization and training methods were the same as in the parameter enhancement system.
	
	\vspace{-2mm}
	
	\subsection{Objective and subjective evaluation results}
	In the objective test, distortions in MELP parameters obtained by clean and enhanced speech signals were evaluated.
	The metrics for measuring distortion were the error rate of the voicing flag at the first frequency band (VUV error; \%), the root mean square error (RMSE) for gain features (Gain-RMSE), the RMSE for F0 (F0-RMSE; Hz), and the log-spectral distance of LSFs in dB (LSD; dB). 
	In addition, the short-time objective intelligibility (STOI) \cite{Taal2010short} was measured to evaluate the intelligibility of reconstructed speech.
	In the evaluation of F0-RMSE, only voiced regions were evaluated.
	Moreover, we also included the codec outputs of {\textit{clean}} and {\textit{noisy}} speech signals to simulate the performance of communication systems with no background noise and no enhancement module, respectively.
	Note that these systems represent the upper and lower bounds of performances, respectively.
	
	The objective results are summarized in Table~\ref{table:obj_results} and \ref{table:stoi_results}.
	Experimental results were as follows.
	(1) All enhancement systems showed significant performance improvement in all metrics (noisy vs. all enhancement systems).
	(2) For the modeling of speech's statistical characteristics such as prosody or voice color, the proposed parameter enhancement showed much higher accuracy than the conventional IRM-based speech enhancement (IRM vs. Param in VUV error, Gain-RMSE, F0-RMSE, and LSD).
	This is because the parameter enhancement is designed to estimate the statistical characteristics of clean speech directly.
	(3) The intelligibility of proposed parameter enhancement was slightly worse than the IRM-based enhancement (STOI).
	Note that the parameter enhancement operates on the vocoder parameter domain, so the cross-correlation between clean and enhanced speech (that is, STOI) is more easily weakened than in IRM-based enhancement, which operates on the frequency domain directly. 
	(4) The decoder side parameter enhancement performed slightly worse than the encoder side parameter enhancement (Param-Enc vs. Param-Dec).
	This implies that the degradation by quantization effect before the DL network is larger than that after the DL network.
	However, their difference was negligible, and the trend with respect to the IRM-based enhancement was the same.
	(5) Although the performance of smaller network was worse than that of the larger one (large network vs. small network), the performance degradation in the proposed parametric approach was much smaller than the IRM-based approach.

	\begin{table}[]
		\caption{Subjective preference test results (\%) between the speech samples.
				The systems that achieved significantly better preference at the $p < 0.01$ level are in bold typeface.}
		\label{table:abx_results}
		\setlength\tabcolsep{4pt} 
		\centering
		\footnotesize
		\begin{tabular}{c||c|c|c|c|c||c}
			\Xhline{2\arrayrulewidth}
			\multirow{2}{*}{\begin{tabular}[c]{@{}c@{}}Test\\ index\end{tabular}} & \multirow{2}{*}{Noisy} & \multirow{2}{*}{IRM} & \multicolumn{2}{c|}{Param} & \multirow{2}{*}{Neutral} & \multirow{2}{*}{$p$-value} \\ \cline{4-5}
			&                        &                      & Enc          & Dec         &                          &                          \\ \hline\hline
			Test 1               & \textbf{7.5}                    & \textbf{89.6}              & --           & --          & \textbf{2.9}                     & $\mathbf{<10^{-124}}$              \\ 
			Test 2               & \textbf{11.3}                   & --                   & \textbf{86.3}         & --          & \textbf{2.5}                      & $\mathbf{<10^{-90}}$              \\ 
			Test 3               & --                     & 21.0                 & 17.3         & --          & 61.7                     & $0.18$                     \\ 
			Test 4               & --                     & --                   & 20.8         & 19.4        & 59.8                     & $1.00$                   \\ \Xhline{2\arrayrulewidth}
		\end{tabular}
		\vspace{-5mm}
	\end{table}
	To evaluate the perceptual quality of the proposed system, the A-B preference test was performed.
	In detail, the randomly selected 20 reconstructed utterances from the test set were mixed with the 7 dB SNR of babble and volvo noise, then enhanced by the small network setup for simulating a real communication environment.
	In the evaluation, total 12 listeners were asked to rate the quality preference and instructed to pay attention to both the signal distortion and the noise intrusiveness.
	The preference test results summarized in Table~\ref{table:abx_results} first verify the significant quality improvement in the MELP coding system using parameter enhancement in addition to the IRM enhancement (Test 1 and 2).
	Moreover, the results shown that the perceptual qualities of all enhancement systems are indistinguishable (Test 3 and 4).
	
	\vspace{-2mm}
		
	\subsection{Computational efficiency of enhancement systems}
	To evaluate the computational efficiency of the systems, we computed the floating point operation per second (FLOPs) of each system.
	For the FLOP of DL network, the operation for biases and activations were neglected, and for the FLOP of the Fourier transform, we assumed the real split-radix fast Fourier transform algorithm \cite{Sorensen1987real}, whose FLOP is 3,078 for a 256-points Fourier transform.
	As a result, we obtained a total computational efficiency around \textbf{5.06 MFLOPs} and \textbf{4.11 MFLOPs} in the small IRM enhancement and parameter enhancement systems, respectively.
	As to the equivalent performance of the IRM and parameter enhancement systems, the proposed parameter enhancement system showed faster computational efficiency than the conventional IRM enhancement system, about 1 MFLOPs.
	\vspace{-2mm}
	
	\section{Conclusion}
	\label{sec:conclusion}	
	In this paper, we introduced a DL-based parameter enhancement method for a MELP speech codec in noisy communication environments. 
	By directly enhancing the MELP parameters, the proposed algorithm was successfully combined with the MELP-based speech communication system.
	Experimental results showed that the proposed method had a higher statistical modeling accuracy in terms of prosody and voice characteristics with faster enhancement speed than the conventional speech enhancement system, while the perceptual quality was similar. 
	In summary, the proposed system successfully constructed a simple and compact speech enhancement system for a low profile speech codec in noisy environments by removing additional processing pipelines.
	\vspace{-2mm}
	\section{Acknowledgment}
	This research was supported by Basic Science Research Program through the National Research Foundation of Korea (NRF) funded by the Ministry of Science and ICT (2019-11-0124).
	
	\vfill
	\pagebreak
	
	\bibliographystyle{IEEEtran}
	
	\bibliography{mybib_mj}
	
\end{document}